\documentclass[twocolumn]{aastex62}

\newcommand{\G}{\textit{Gaia}}
\newcommand{\DR}{\textit{Gaia}~DR2}
\newcommand{\Sp}{\textit{Spitzer}}
\newcommand{\CW}{CatWISE}
\graphicspath{{./}{figures/}}

\received{December 4, 2019}
\revised{December 10, 2019}
\accepted{December 14, 2019}
\submitjournal{ApJ Letters}

\usepackage{bm}
\usepackage{xfrac}

\shorttitle{CWISEP J1446--2317}
\shortauthors{Marocco et al.}

\begin{document}

\title{Improved infrared photometry and a preliminary parallax measurement for the extremely cold brown dwarf CWISEP~J144606.62$-$231717.8}


\correspondingauthor{Federico Marocco}
\email{federico.marocco@jpl.nasa.gov}

\author[0000-0001-7519-1700]{Federico Marocco}\altaffiliation{NASA Postdoctoral Program Fellow}
\affiliation{Jet Propulsion Laboratory, California Institute of Technology, 4800 Oak Grove Dr., Pasadena, CA 91109, USA}
\affiliation{IPAC, Mail Code 100-22, Caltech, 1200 E. California Blvd., Pasadena, CA 91125, USA}

\author[0000-0003-4269-260X]{J. Davy Kirkpatrick}
\affiliation{IPAC, Mail Code 100-22, Caltech, 1200 E. California Blvd., Pasadena, CA 91125, USA}

\author[0000-0002-1125-7384]{Aaron M. Meisner}
\affiliation{NSF's National
Optical-Infrared Astronomy Research Laboratory, 950 N. Cherry Ave., Tucson, AZ 85719, USA}

\author[0000-0001-7896-5791]{Dan Caselden}
\affiliation{Gigamon Applied Threat Research, 619 Western Avenue, Suite 200, Seattle, WA 98104, USA}

\author{Peter R. M. Eisenhardt}
\affiliation{Jet Propulsion Laboratory, California Institute of Technology, 4800 Oak Grove Dr., Pasadena, CA 91109, USA}

\author[0000-0001-7780-3352]{Michael C. Cushing}
\affiliation{Department of Physics and Astronomy, University of Toledo, 2801 West Bancroft St., Toledo, OH 43606, USA}

\author[0000-0001-6251-0573]{Jacqueline K. Faherty}
\affiliation{Department of Astrophysics, American Museum of Natural History, Central Park West at 79th Street, NY 10024, USA}

\author{Christopher R. Gelino}
\affiliation{IPAC, Mail Code 100-22, Caltech, 1200 E. California Blvd., Pasadena, CA 91125, USA}

\author[0000-0001-5058-1593]{Edward L. Wright}
\affiliation{Department of Physics and Astronomy, UCLA, 430 Portola Plaza, Box 951547, Los Angeles, CA 90095-1547, USA}


\begin{abstract}
We present follow-up \Sp\ observations at 3.6$\mu$m (ch1) and 4.5$\mu$m (ch2) of CWISEP~J144606.62--231717.8, one of the coldest known brown dwarfs in the solar neighborhood. This object was found by mining the Wide-field Infrared Survey Explorer (\textit{WISE}) and \textit{NEOWISE} data via the \CW\ Preliminary Catalog by \citet{2019arXiv191112372M}, where an initial \Sp\ color of ch1--ch2 = 3.71$\pm$0.44\,mag was reported, implying it could be one of the reddest, and hence coldest, known brown dwarfs. Additional \Sp\ data presented here allows us to revise its color to ch1--ch2 = 2.986$\pm$0.048\,mag, which makes CWISEP~J144606.62--231717.8 the 5th reddest brown dwarf ever observed. A preliminary trigonometric parallax measurement, based on a combination of \textit{WISE} and \Sp\ astrometry, places this object at a distance of 10.1$^{+1.7}_{-1.3}$\,pc. Based on our improved \Sp\ color and preliminary parallax, CWISEP~J144606.62--231717.8 has a $T_{\rm eff}$ in the 310--360\,K range. Assuming an age of 0.5--13\,Gyr, this corresponds to a mass between 2 and 20\,$M_{\rm Jup}$.
\end{abstract}




\section{Introduction} 
\label{sec:intro}
Ever since its discovery in 2014, WISE~J085510.83--071442.5 \citep[hereafter W0855]{2014ApJ...786L..18L} has remained the coldest brown dwarf known. With an estimated effective temperature of $\sim250$\,K, W0855 represents an isolated extreme of the substellar spectral sequence. The census of the coldest, lowest mass constituents of the solar neighborhood is however known to be incomplete. \citet{2019ApJS..240...19K} have estimated the current completeness limit to be 19\,pc in the 900--1050\,K interval, but to decrease to only 8\,pc in the 300--450\,K interval. At even lower $T_{\rm eff}$, W0855 is the only object known.

Obtaining a more complete census of extremely cold brown dwarfs is a fundamental step towards robustly constraining the efficiency and history of the star formation process at its lowest mass \citep{2019ApJS..240...19K}. Solivagant objects with mass as low as a few Jupiter masses ($M_{\rm Jup}$) have been found in star formation regions and nearby, young moving groups \citep{2019AJ....158...54E,2018AJ....156...76L,2018MNRAS.473.2020L,2017ApJ...842...65Z,2016ApJS..225...10F}. Older, isolated objects with these masses therefore should exist, and numerical simulations show that their space density is extremely sensitive to the low-mass cutoff for star formation \citep{2019ApJS..240...19K}.
 
Using data from the recently released \CW\ Preliminary Catalog \citep{2019arXiv190808902E} and a combination of machine learning and color-, magnitude-, and proper motion-based selection criteria, \citet[hereafter M19]{2019arXiv191112372M} identified a large sample of candidate cool brown dwarfs in the solar neighborhood. Through a dedicated \Sp\ observing campaign to obtain 3.6$\mu$m (ch1) and 4.5$\mu$m (ch2) data and improved proper motion measurement, M19 confirmed 114 objects in their sample to be nearby brown dwarfs, with 17 of them having \Sp\ ch1--ch2 color clearly indicating $T_{\rm eff} < 460$\,K, corresponding to spectral type Y0 or later. CWISEP~J144606.62--231717.8 (hereafter CW1446) stands out among them, with a ch1--ch2 color of 3.71$\pm$0.44\,mag, potentially supplanting W0855 (ch1--ch2=$3.55\pm0.07$\,mag) as the reddest and therefore coldest brown dwarf known. Here we present additional \Sp\ observations that better constrain the color of this source, and provide a preliminary measurement of its parallax.

In Section~\ref{sec:selection}, we briefly summarize the brown dwarf candidate selection that led to the discovery of CW1446 and the data available prior to this paper. In Section~\ref{sec:spitzer} we present new \Sp\ follow-up observations and the resulting improved photometry, and in Section~\ref{sec:astrometry} we combine all of the \Sp\ and \textit{WISE} astrometry to obtain a preliminary parallax measurement. In Section~\ref{sec:analysis} we derive the basic properties for CW1446, and in Section~\ref{sec:discussion} we put this new object into context and discuss future work.

\section{Source selection and existing data}
\label{sec:selection}
CW1446 was found as part of our larger effort to complete the census of very cold brown dwarfs in the solar neighborhood using the \CW\ Preliminary Catalog\footnote{available at \url{https://irsa.ipac.caltech.edu/cgi-bin/Gator/nph-scan?mission=irsa&submit=Select&projshort=WISE} and \url{catwise.github.io}}, an infrared photometric and astrometric catalog consisting of 900,849,014 sources over the entire sky selected from {\it WISE} and {\it NEOWISE} data collected from 2010 to 2016 at $W1$ (3.4$\mu$m) and $W2$ (4.6$\mu$m) \citep{2019arXiv190808902E}.

The search was conducted using the \textsc{Python} package \textit{XGBoost}\footnote{\url{https://xgboost.readthedocs.io/en/latest/}} \citep{Chen:2016:XST:2939672.2939785}, which implements machine learning algorithms under the gradient boosting framework. A detailed description of the search procedure is given in \citet{2019ApJ...881...17M}, and here we only briefly summarise the most important steps.

We trained the \textit{XGBoost} model with a set of known T and Y dwarfs taken from the literature, cross-matched against \CW\ to obtain their \CW\ data. The model was trained on a set of the \CW\ data available for a given source, including aperture and PSF photometry, proper motion, the $\chi^2$ of the measurements, and artifact flags. Sample weights were applied to mitigate the class imbalance in the training set.

After training the \textit{XGBoost} classifier with our initial training set, we applied it to the entire \CW\ catalog, and selected $\sim$10,000 objects with the highest predicted probability of being cold brown dwarfs. We then visually inspected each object, using available optical, near- and mid-infrared images and the online image blinking/visualization tool WiseView\footnote{\url{http://byw.tools/wiseview}} \citep{2018ascl.soft06004C}. Objects passing this inspection, with $W1-W2$ color visually consistent with $W1-W2 > 1$\,mag, and with by-eye motion\footnote{Roughly \sfrac{1}{2} pixel over the 8-years baseline, or $\approx170$\,mas\,yr$^{-1}$}, were added to the training set. We then iterated by re-training the classifier on the full training data, and applied the re-trained classifier to the entire catalog to select another batch of high probability positive class entries.  
The selection yielded an initial sample of 32 late-T and Y dwarf candidates, with either no detection or a marginal detection in W1 and visible motion. These were followed up through our \Sp\ campaign (program 14034, Meisner, PI) to obtain ch1 and ch2 photometry to estimate effective temperature and photometric distance. The results are presented in M19. CW1446 is the reddest (therefore coldest) among the objects presented in M19, with ch2 = $15.802\pm0.024$\,mag, and ch1--ch2 = 3.71$\pm$0.44\,mag. Photometric data available prior to our follow-up is summarized in Table~\ref{tab:photometry}. 

\section{{\it Spitzer} follow-up}
\label{sec:spitzer}
\begin{figure*}
    \centering
    \includegraphics[width=\textwidth, trim={3cm 0 3cm 0}, clip]{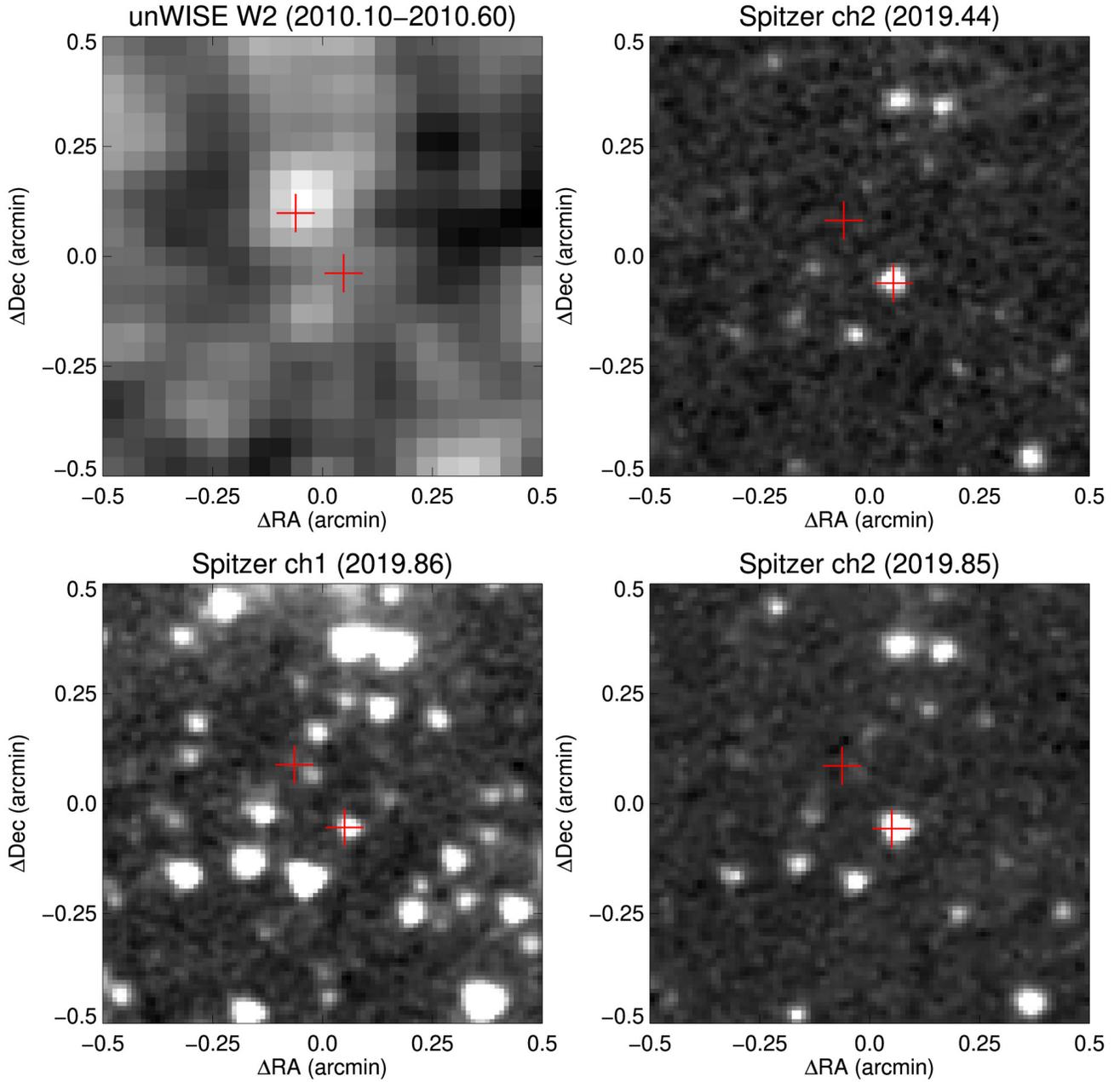}
    \caption{$1\times1$ arcmin cutouts from the unWISE W2 epoch coadd \citep[top left][]{2019PASP..131l4504M}, and the \Sp\ ch1 and ch2 mosaics, centered around CW1446. Red crosses mark its position at the earliest unWISE epoch (2010.10--2010.60), and the first \Sp\ epoch (2019.44). The second \Sp\ epoch exhibits motion along the R.A. axis which is not consistent with the proper motion of the source, hinting at its large parallactic motion (see Section~\ref{sec:astrometry} for details).}
    \label{fig:images}
\end{figure*}

Follow-up \Sp\ ch1 observations were taken as part of program 14307 (Marocco, PI). We took thirty-six exposures of 100\,s, using a random dither pattern of medium scale. The total integration time was designed to achieve SNR$\sim$10, based on the ch1 magnitude from our PID 14034 data. Photometry was measured following the same procedure described in \citet{2019ApJ...881...17M}. The new ch1 mosaic is presented in Figure~\ref{fig:images}.

\begin{table*}
    \centering
    \caption{Photometry and astrometry for CW1446.}
    \label{tab:photometry}
    \begin{tabular}{l c c c c}
    \hline
    Parameter & Units & Value & Ref. & Notes \\
    \hline
    FLAMINGOS-2 $J$ & mag & $>$22.36 & M19 & \\
    CatWISE W1 &  mag & 18.281$\pm$0.292 & M19 & motion fit \\
    CatWISE W2 & mag & 15.998$\pm$0.094 & M19 & motion fit \\
    \Sp\ ch1 & mag & 19.682$\pm$0.424 & M19 & aperture -- May 2019 \\
    \Sp\ ch2 & mag & 15.915$\pm$0.022 & M19 & aperture -- May 2019 \\
    \Sp\ ch1 & mag & 19.340$\pm$0.445 & M19 & PRF fit -- May 2019 \\
    \Sp\ ch2 & mag & 15.689$\pm$0.026 & M19 & PRF fit -- May 2019 \\
    \Sp\ ch1 & mag & 18.951$\pm$0.034 & this letter & aperture -- Nov. 2019 \\
    \Sp\ ch2 & mag & 15.927$\pm$0.017 & this letter & aperture -- Nov. 2019\\
    \Sp\ ch1 & mag & 18.905$\pm$0.045 & this letter & PRF fit -- Nov. 2019\\
    \Sp\ ch2 & mag & 15.919$\pm$0.018 & this letter & PRF fit -- Nov. 2019\\
    $\varpi$ & mas & 99.2$\pm$14.7 & this letter & \\
    $\mu_\alpha \cos \delta$ & mas yr$^{-1}$ & --794.3$\pm$51.9 & this letter & \\
    $\mu_\delta$ & mas yr$^{-1}$ & --964.8$\pm$30.7 & this letter & \\
    v$_{\rm tan}$ & km s$^{-1}$ & 59.7$\pm$9.0 & this letter & \\
    \hline
    \end{tabular}
\end{table*}

We also obtained \Sp\ ch2 photometry as part of program 14224 (Kirkpatrick, PI). Because these observations were intended for high-precision astrometry, we designed the observations to have SNR $>$ 100 at each epoch. Given the brightness of CW1446 (W2$\sim$15.8\,mag), we took 9 exposures of 100s using a random dither pattern of medium scale.

We performed both aperture and PRF-fit photometric measurements using using the \Sp\ MOsaicker and Point Source EXtractor with point-source extraction package \citep[MOPEX/APEX;][]{2005ASPC..347...81M,2005PASP..117.1113M}. Custom mosaics were built to provide better cosmic-ray removal than the default post basic calibrated data files (pBCD) provide. For this custom processing we coadded the corrected basic calibrated data (CBCD) frames and ran detections on the resultant coadd. Raw fluxes were then measured by MOPEX/APEX using the stack of individual CBCD files that comprised the coadd. These raw fluxes were converted to magnitudes by applying aperture corrections and comparing to the published ch1 and ch2 flux zero points, as described in section 5.1 of \citet{2019ApJS..240...19K}.

The new ch1 and ch2 measurements, presented in Table~\ref{tab:photometry}, yield a revised ch1--ch2 color of 2.986$\pm$0.048\,mag (PRF; the aperture color is 3.024$\pm$0.038\,mag). The new color is significantly bluer than its preliminary value (3.71$\pm$0.44\,mag), mostly because of the large difference in ch1. The measured ch1 PRF flux from the early \Sp\ observations is 5.3$\pm$2.1\,$\mu$Jy, while the new measurement is 7.86$\pm$0.31\,$\mu$Jy, corresponding to a 1.2$\sigma$ difference, while the aperture flux measurements are 3.1$\pm$1.2\,$\mu$Jy and 6.11$\pm$0.18\,$\mu$Jy respectively, corresponding to a 2.5$\sigma$ difference.

\section{Astrometry}
\label{sec:astrometry}

\citet{2019ApJS..240...19K} describes the methodology used to measure astrometry from the \Sp\ images, but we have made a few improvements since then. First, we now match bright re-registration stars in each frame to \DR\ and use only those \G\ stars that have full five-parameter solutions. Second, in order to assure that we have enough stars per frame with which to do the re-registration, we select stars down to a SNR value of 30. For CW1446, this resulted in 56 re-registration stars. Third, because we have chosen re-registration stars with full astrometric solutions, we can predict their absolute positions at the time of each \Sp\ epochal observation, thus allowing us to measure astrometry on the absolute reference frame from the start. No relative-to-absolute adjustment is therefore needed.

The original \Sp\ ch2 observation from program 14034 (Meisner, PI) was the only one obtained in the penultimate observing window. We have requested six additional ch2 observations in the final \Sp\ observing window, which was open from 2019 early-November through mid-December. We present the first of those six observations here. These \Sp\ data alone, however, are not sufficient to decouple proper motion and parallax. For this we relied on \textit{WISE} W2 detections. Specifically, we took the twelve unWISE epochal coadds \citep[and references therein]{2019PASP..131l4504M} spanning the range 2010 February to 2018 July, and performed \textsc{crowdsource} \citep{2019ApJS..240...30S,2018ApJS..234...39S} detections on the full unWISE tile containing the position of CW1446 (tile 2215m228, centered on R.A.=221.5$^\circ$, Dec.=--22.8$^\circ$). For each epoch, we matched these detections to objects in \DR\ with full five-parameter solutions. These \G\ objects were placed at their expected positions at the time of the \textit{WISE} observations so that, again, astrometry could be re-registered onto the absolute \DR\ reference frame. Additional information on this process can be found in M19. These unWISE data were then associated with the position of the Earth at the mean time of each unWISE epoch, and an astrometric fit was run using the prescription discussed in Section 5.2.3 of \citet{2019ApJS..240...19K}. 

\begin{figure*}
    \centering
    \includegraphics[width=\textwidth,trim={0 9.6cm 0 0.7cm}, clip]{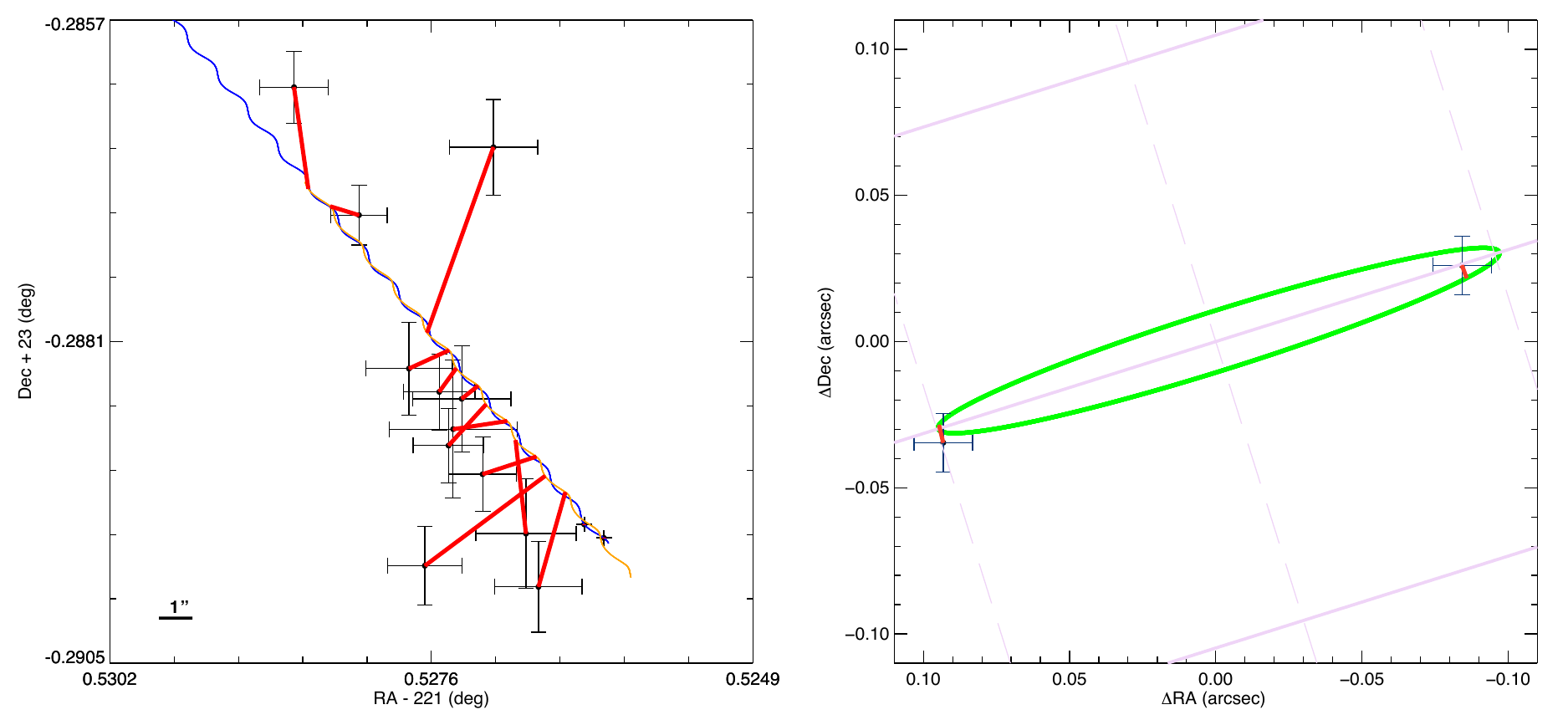}
    \caption{Proper motion + parallax fit to the combined \Sp\ and unWISE W2 data for CW1446. \textit{Left}: The full astrometric solution and full set of empirical measurements. The unWISE W2 epochal data are shown by the black points with large error bars and the \Sp\ data are shown by the the black points with the much smaller error bars. The fit of the astrometric path of the object as seen from \Sp\ is shown by the blue curve, and the astrometric path as seen from the Earth is shown by the orange curve. The red lines connect each data point with the spot on the relevant curve at that epoch. \textit{Right}: A square patch of sky centered at the mean equatorial position of the target. The green curve is the parallactic fit, which is just the blue curve in the previous panel with the proper motion vector removed. Solid and dashed pale purple lines are the ecliptic latitude and longitude coordinate grid, respectively. This panel omits, for clarity, the less accurate unWISE astrometry.}
    \label{fig:pm}
\end{figure*}

The resulting fit is given in Table~\ref{tab:photometry} and illustrated in Figure~\ref{fig:pm}. The parallactic solution should be considered preliminary and of low confidence because there is only a single high-quality data point anchoring each side of the parallactic ellipse. The low confidence of the solution is also reflected in the large parallactic error of $\sim$15\%.

\section{Analysis}
\label{sec:analysis}
\begin{figure*}
    \centering
    \includegraphics[width=0.49\textwidth, trim={1.9cm 3cm 1.5cm 3cm}, clip]{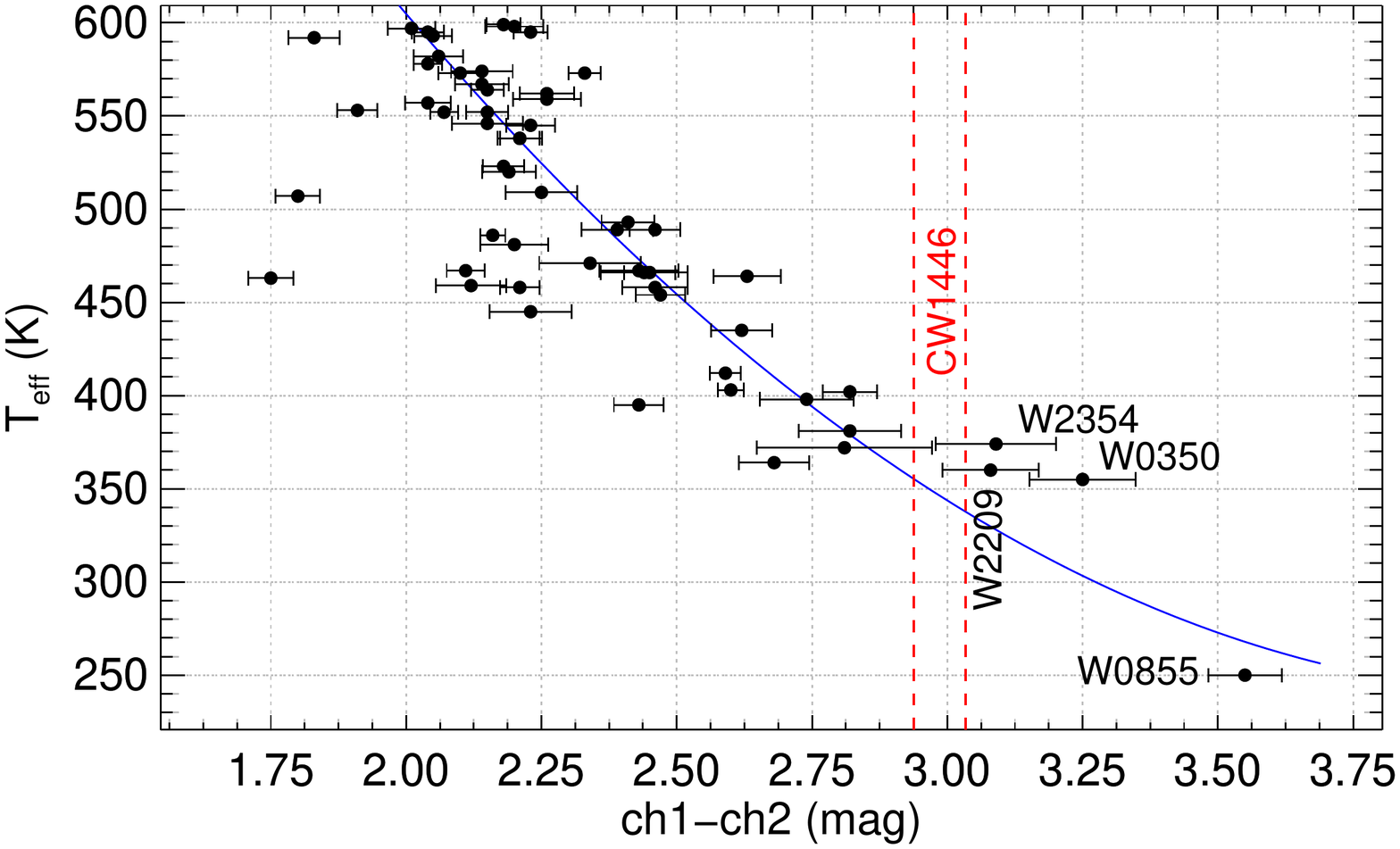}
    \includegraphics[width=0.49\textwidth, trim={1.9cm 3cm 1.5cm 3cm}, clip]{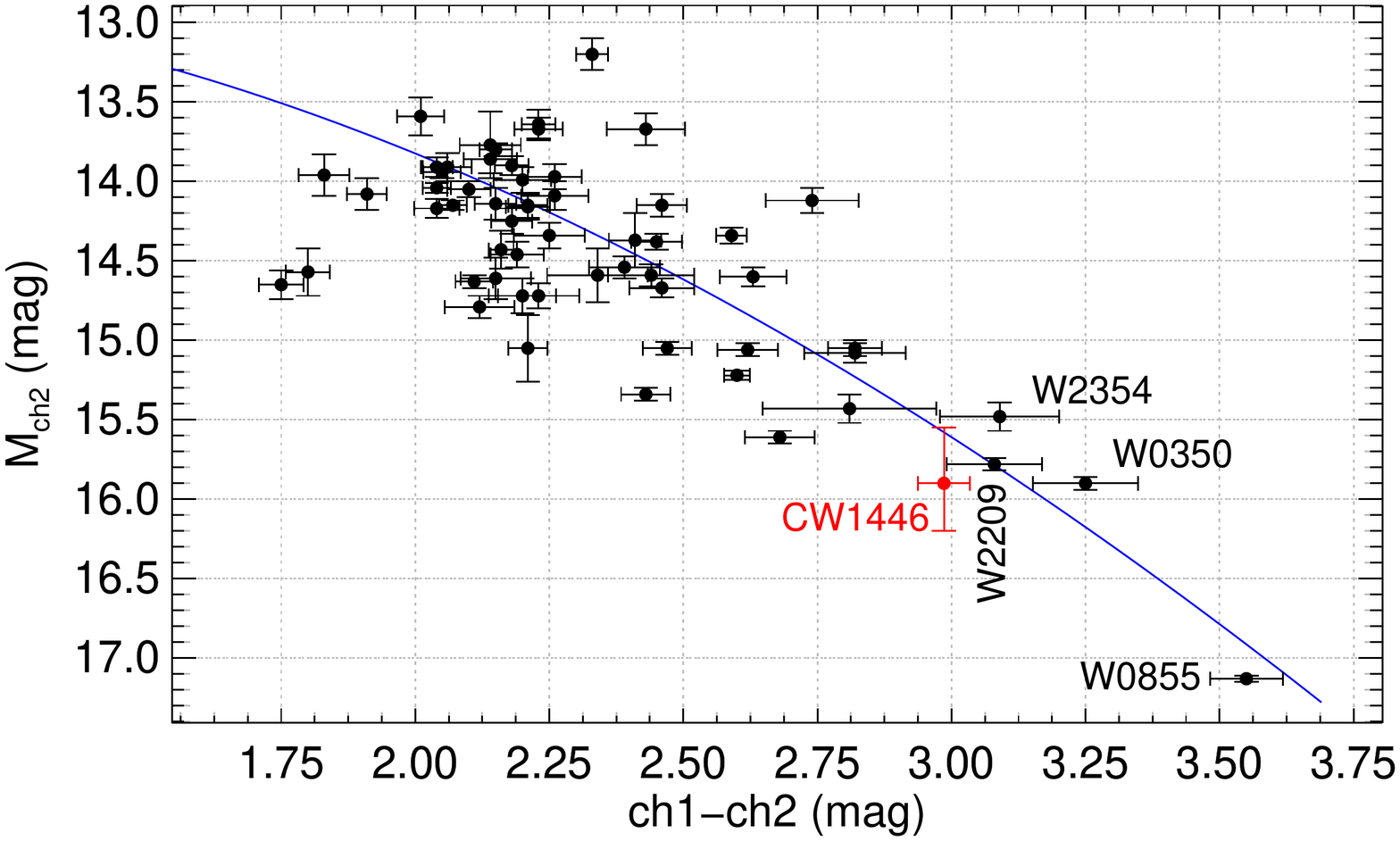}
    \caption{$T_{\rm eff}$ (left panel) and absolute \Sp\ ch2 magnitude (right panel) as a function of \Sp\ ch1--ch2 colors for nearby late-T and Y dwarfs. Black points are all objects with $T_{\rm eff} < 600$\,K and measured parallaxes taken from \citet[Table 8]{2019ApJS..240...19K}. The red dashed lines in the left panel encompass the $1\sigma$ color range for CW1446. Overplotted in blue is the polynomial relation presented in \citet{2019ApJS..240...19K}. The four objects redder than CW1446 are WISE~J220905.73+271143.9 \citep[labelled W2209 on the plot;][]{2011ApJS..197...19K}, WISEA~J235402.79+024014.1 \citep[W2354;][]{2015ApJ...804...92S}, WISE~J035000.32--565830.2 \citep[W0350;][]{2012ApJ...753..156K}, and WISE~J085510.83--071442.5 \citep[W0855;][]{2014ApJ...786L..18L}.}
    \label{fig:cmd}
\end{figure*}

With a ch1--ch2 color of 2.986$\pm$0.048, and a distance of 10.1$^{+1.7}_{-1.3}$\,pc, CW1446 is the one of the reddest, least luminous, and therefore likely coldest brown dwarfs known in the solar neighborhood. Figure~\ref{fig:cmd} shows $T_{\rm eff}$ and $M_{\rm ch2}$ as a function of \Sp\ ch1--ch2 color for a sample of known late-T and Y dwarfs from the literature \citep[see][and references therein]{2019ApJS..240...19K}. The ch1--ch2 to $T_{\rm eff}$ and M$_{\rm ch2}$ to $T_{\rm eff}$ polynomial relations presented in \citet{2019ApJS..240...19K} imply a $T_{\rm eff}$ in the range $\sim$310--360\,K for CW1446 (see Figure~\ref{fig:cmd}).

For such a cold $T_{\rm eff}$, and if we assume CW1446 is a field object (i.e. with age in the $\sim$500\,Myr -- 13\,Gyr range), the BT-Settl models \citep{2012EAS....57....3A,2013MmSAI..84.1053A} imply a mass in the range 2--20\,$M_{\rm Jup}$. However, given its relatively high tangential velocity (59.7$\pm$9.0\,km\,s$^{-1}$), CW1446 is unlikely to be very young. If we assume CW1446 is coeval with the population of nearby ultracool dwarfs, whose age is in the range $\sim$1.5--6.5\,Gyr \citep[see e.g.][and references therein]{2018PASP..130f4402W}, we find its mass to be between 4 and 14\,$M_{\rm Jup}$. 

Despite being slightly bluer, our preliminary parallax suggests CW1446 is as luminous as WISE~J035000.32--565830.2, currently the second reddest brown dwarf known (ch1--ch2=3.25$\pm$0.10\,mag, M$_{\rm ch2}=15.90\pm0.04$\,mag). Comparison to the Y0 dwarf spectral standard, WISE~J173835.53+273259.0 \citep{2011ApJ...743...50C}, shows that CW1446 is clearly redder (ch1--ch2$=2.986\pm0.048$\,mag vs. $2.620\pm0.056$\,mag) and less luminous (M$_{\rm ch2}=15.90\pm0.04$\,mag vs. $15.06\pm0.04$\,mag). Interpolating the spectral type to \Sp\ color and M$_{\rm ch2}$ relations presented in \citet{2019ApJS..240...19K}, we find CW1446 would have a spectral type of $\approx$Y1.5. However, we warn the reader that the scatter in the spectral type to color and magnitude relations for such cold objects is still not well quantified or understood, with spectroscopically classified Y0 dwarfs occupying a $\sim$1\,mag range in ch1--ch2 and a $\sim1.3$\,mag range in M$_{\rm ch2}$ \citep[see Figure 4 and 5 in][]{2019ApJS..240...19K}. Moreover, the \Sp\ ch1 and ch2 photometry probes a different wavelength regime than the near-infrared spectral types, which are defined based on the morphology of the J- and H-band spectra \citep{2011ApJ...743...50C}, and are therefore likely sensitive to different physical and chemical processes. Therefore further interpretation of CW1446 with respect to the rest of the cold brown dwarf population based on \Sp\ data alone is unwarranted.

Shorter wavelength photometric detections are unavailable for this object, given that it is well below the detection threshold for existing optical and near-infrared surveys. Our dedicated FLAMINGOS-2 observations lead to MKO $J > 22.36$\,mag (M19), implying $J-{\rm ch2} > 6.44$\,mag, consistent with the $T_{\rm eff}$ derived here. 

\section{Discussion}
\label{sec:discussion}
The upcoming \Sp\ astrometric observations will allow us to improve the constraint on the distance to this object, securing one of the two vertices of the parallactic ellipse. However, due to the end of the \Sp\ mission, no further measurement is possible at the opposite vertex, limiting the improvement we can expect. Further characterization of CW1446 anyway requires spectroscopic follow-up. Given the $T_{\rm eff}$ estimate and preliminary distance measurement presented here, the expected $H$ magnitude for CW1446 is 24--25.5 mag, a depth prohibitive for ground-based spectroscopy with existing facilities. Spectroscopic characterization can therefore only be provided by the upcoming \textit{James Webb Space Telescope.}

CW1446 occupies the sparsely populated $300-400$\,K regime. W0855 however remains the only $T_{\rm eff} < 300$\,K object known to date. Given the brightness and proximity of W0855, \citet{2014AJ....148...82W} estimated that the existing \textit{WISE} data should contain of order 4--35 ``W0855-like'' objects, and predicted that, if such objects did indeed exist, astrometric analysis of the combination of AllWISE and \textit{NEOWISE} data would allow their discovery. 

Yet W0855-like objects remain elusive, despite investigations of \textit{WISE} and \textit{NEOWISE} data using the \CW\ Preliminary Catalog \citep{2019arXiv190808902E} and the ``Backyard Worlds: Planet 9'' citizen science project \citep{2017ApJ...841L..19K}.

The upcoming \CW\ 2020 catalog will be based on the full set of publicly available \textit{WISE} and \textit{NEOWISE} data covering the 2010--2018 baseline, and achieves significantly better completeness and motion sensitivity, so may reveal colder objects. Further advancement on the question of whether there is a low mass cutoff to star formation may need to wait for the Near Earth Object Surveyor (formerly NEOCam), which will provide even deeper imaging of most of the sky at wavelengths similar to W2, with a mission length of at least 5 years. 

\acknowledgments
This research was partly carried out at the Jet Propulsion Laboratory, California Institute of Technology, under a contract with NASA. 

FM is supported by an appointment to the NASA Postdoctoral Program at the Jet Propulsion Laboratory, administered by Universities Space Research Association under contract with NASA.

AMM acknowledges support from Hubble Fellowship HST-HF2-51415.001-A.

CatWISE is funded by NASA under Proposal No. 16-ADAP16-0077 issued through the Astrophysics Data Analysis Program, and uses data from the NASA-funded WISE and NEOWISE projects.

This work is based in part on observations made with the Spitzer Space Telescope, which is operated by the Jet Propulsion Laboratory, California Institute of Technology under a contract with NASA.

\facilities{Spitzer (IRAC), \textit{WISE/NEOWISE}}
\software{\textit{XGBoost} \citep{Chen:2016:XST:2939672.2939785}, MOPEX \citep{2005ASPC..347...81M,2005PASP..117.1113M}, WiseView \citep{2018ascl.soft06004C}}

\bibliographystyle{aasjournal}
\bibliography{refs.bib}

\end{document}